\documentclass[prd,twocolumn,showpacs,amsmath,amssymb,superscriptaddress,floatfix,nofootinbib]{revtex4}

\usepackage{amsfonts}
\usepackage{color}
\usepackage{graphicx,amsfonts,multirow}
\usepackage{amsmath}
\usepackage{url}

\usepackage{ulem}

\begin{document}


\title{The $B\to \pi \ell\nu_\ell$ semileptonic decay within the LCSR approach under heavy quark effective field theory}

\author{Rui-Yu Zhou}
\email{zhoury@cqu.edu.cn}
\address{Department of Physics, Chongqing University, Chongqing 401331, People's Republic of China}

\author{Lei Guo}
\email{guoleicqu@cqu.edu.cn}
\address{Department of Physics, Chongqing University, Chongqing 401331, People's Republic of China}

\author{Hai-Bing Fu}
\email{fuhb@cqu.edu.cn}
\address{Department of Physics, Guizhou Minzu University, Guiyang 550025, People's Republic of China}

\author{Wei Cheng}
\email{chengw@cqu.edu.cn}
\address{Department of Physics, Chongqing University, Chongqing 401331, People's Republic of China}

\author{Xing-Gang Wu}
\email{wuxg@cqu.edu.cn}
\address{Department of Physics, Chongqing University, Chongqing 401331, People's Republic of China}

\date{\today}

\begin{abstract}

The heavy quark effective field theory (HQEFT) provides an effective way to deal with the heavy meson decays. In the paper, we adopt two different correlators to derive the light-cone sum rules of the $B \to \pi$ transition form factors (TFFs) within the framework of HQEFT. We label those two LCSR results as LCSR-${\cal U}$ and LCSR-${\cal R}$, which are for conventional correlator and right-handed correlator, respectively. We observe that the correlation parameter $|\rho_{\rm RU}|$ for the branching ratio ${\cal B}(B \to \pi l \nu_{l})$ is $\sim 0.85$, implying the consistency of the LCSRs under different correlators. Moreover, we obtain $|V_{\rm ub}| _{{\rm LCSR}-{\cal U}}=(3.45^{+0.28}_{-0.20}\pm{0.13}_{\rm{exp}})\times10^{-3}$ and $|V_{\rm ub}| _{{\rm LCSR}-{\cal R}} =(3.38^{+0.22}_{-0.16} \pm{0.12}_{\rm{exp}})\times10^{-3}$. We then obtain $\mathcal{R}_{\pi}| _{{\rm LCSR}-{\cal U}}=0.68^{+0.10}_{-0.09}$ and $\mathcal{R}_{\pi}| _{{\rm LCSR}-{\cal R}}=0.65^{+0.13}_{-0.11}$, both of them agree with the Lattice QCD predictions. Thus the HQEFT provides a useful framework for studying the $B$ meson decays. Moreover, by using right-handed correlator, the twist-2 terms shall dominant the TFF $f^+(q^2)$, which approaches over $\sim97\%$ contribution in the whole $q^2$-region; and the large twist-3 uncertainty for the conventional correlator is greatly suppressed. One can thus adopt the LCSR-${\cal R}$ prediction to test the properties of the various models for the pion twist-2 distribution amplitudes.

\end{abstract}


\maketitle

\section {Introduction}

The $B$-meson semileptonic decay is an important channel for testing the heavy quark effective field theories. The mass of the $\tau$ lepton is much larger than the masses of the electron and muon, and the $\tau$ lepton is more sensitive to new physics associated with electroweak symmetry breaking. In recent measurements, the BABAR, Belle and LHCb collaborations have observed large disagreement with the Standard Model (SM) predictions on the ratio $\mathcal{R}_{D^{(*)}} = \mathcal{B}(B\to D ^{(*)} \tau \bar{\nu_\tau})/\mathcal{B}(B\to D ^{(*)}\ell \bar{\nu_\ell})$, where $\ell$ stands for the light lepton $e$ or $\mu$. The weighted average of those measurements gives $\mathcal{R}_{D} = 0.407\pm0.039\pm0.024$ and $\mathcal{R}_{D^{*}} = 0.306\pm0.013\pm0.007$~\cite{HFAG_Amhis:2014hma}. The typical SM predictions are $\mathcal{R}_{D} = 0.299\pm0.011$~\cite{LQCD_Lattice:2015rga} and $\mathcal{R}_{D^{(*)}} = 0.252\pm0.003$~\cite{Fajfer:2012vx}. Meanwhile, the LHCb collaboration also measured the similar ratio $\mathcal{R}_{J/\psi} = 0.71\pm0.17\pm0.18$~\cite{Aaij:2017tyk}, which is 2 $\sigma$ deviation from the SM prediction. Those inconsistencies have aroused people's great interests on searching the new physics (NP) beyond the SM from $B$-meson semileptonic decays.

For example, it is helpful to know whether such kind of disagreement also occurs in $B\to \pi$ semi-leptonic decays. Using the Belle data set of $711fb^{-1}$, the Belle Collaboration predicted the signal strength $\mu=1.52\pm{0.72}$~\cite{Hamer:2015jsa}, where $\mu=1$ corresponds to $\mathcal{B}(B\to \pi \tau \bar{\nu_{\tau}})=1.0\times 10^{-4}$. With the Belle measurement data, the Ref.~\cite{Bernlochner:2015mya} indicates the ratio is $\mathcal{R}^{meas}_{\pi}\simeq 1.05\pm{0.51}$. Theoretically, the Lattice QCD calculation of the RBC and UKQCD Collaborations gives $\mathcal{R}_{\pi}= 0.69\pm{0.19}$~\cite{Flynn:2015mha}. Using the Lattice QCD prediction of the $B\to\pi$ transition form factors (TFFs) given by the Fermilab Lattice and MILC collaborations~\cite{Lattice:2015tia}, Ref.\cite{Bernlochner:2015mya} predicted $\mathcal{R}_{\pi}=0.641\pm0.016$, which was alternated to $\mathcal{R}^{\rm 2HDM}_{\pi}=1.01\pm0.04$ by using the type-II two Higgs doublet model (2HDM).

Before confirming the signal of NP, we need more effects to achieve an accurate SM prediction. The ratio $\mathcal{R}_{\pi}$ strongly depends on the non-perturbative $B\to\pi$ transition form factors (TFFs), which have been predicted by various approaches, such as the Lattice QCD approach, the perturbative QCD approach and the QCD light-cone sum rule (LCSR) approach, cf.Refs.~\cite{Lattice:2015tia, Flynn:2015mha, Lu:2000hj, Kurimoto:2001zj, Huang:2005kk, Wang:2012ab, Khodjamirian:2011ub, Bharucha:2012wy, Imsong:2014oqa, Shen:2016hyv, Huang:2001xb, Ball:2004ye, Belyaev:1993wp}. Each approach has its reliable application region, which can be extended to the physically allowable region via proper extrapolations. A consistent analysis can be achieved by properly connecting predictions under different approaches, which are suitable for various $q^2$-region~\cite{Huang:2004hw}. Among them, the LCSR approach is applicable in both small and intermediate $q^2$ regions and provides an important role for studying the properties of $B\to\pi$ TFFs.

The $B$-meson is made up of a heavy $b$ (or $\bar{b}$) quark and one light quark (or antiquark), whose semileptonic decays can be described by using the heavy quark effective theory. As for the effective theory of heavy quark, there are some criteria \cite{Burdman:1993es} to judge whether the heavy quark expansion is valid, such as: $\bar\Lambda/(2m_Q)\ll1$ and $v\cdot p/(2m_Q)\ll1$, where the $\bar\Lambda$ is the effective mass of the light degrees of freedom in the initial heavy meson. While in the B meson, b quark is much heavier than the other component. Therefore the energy of the light degrees of freedom is far less than the twice of the b quark mass, whereas the second ratio varies roughly between 0 and 1/4 for $m_\pi\leq v\cdot p \leq (m_B^2+m_\pi^2)/(2m_B)$. These criteria are satisfied in this process. In addition to the heavy quark effective theory~\cite{Georgi:1990um, Falk:1990yz, Neubert:1993mb}, the heavy quark effective field theory (HQEFT) is also an effective theory which including the effects of the mixing terms between quark and antiquark fields. The HQEFT enables one to separate the long-distance dynamics from the short one via a systematic way~\cite{Wang:1999zd}, which could lessen the number of non-perturbative wave functions or transition form factors to a certain degree and improve the precision of the calculations. The characteristic feature of HQEFT lies in that its Lagrangian has kept the `particle', `anti-particle', `small' and `large' field components. Basing the HQEFT framework, the LCSR approach can be adopted for a consistent description on the heavy to light meson semileptonic decays ~\cite{Wang:2001mi, Wang:2002zba, Wang:2003yx}. In the present paper, we shall give a detailed LCSR analysis of those TFFs within the framework of the HQEFT.

Usually, the current structure of the correlator is constructed by equating its quantum number $J^P$ (total angular momentum $J$ and parity $P$) as that of the $B$ meson. Constructed via this way, the LCSRs for the $B\to\pi$ TFFs can be arranged following the twist structures of pion light-cone distribution amplitudes (LCDAs). However because the high twist LCDAs are still of high uncertainty, those terms involving high-twist LCDAs provide one of the main uncertainties to the LCSRs of the $B\to \pi$ TFFs. It has been pointed out that by choosing a proper correlator, e.g. by using a chiral current, one can suppress the uncertainties from the high-twist LCDAs and achieve a more accurate LCSR prediction on the TFFs~\cite{Huang:2001xb, Wan:2002hz, Zuo:2006dk, Wu:2007vi, Wu:2009kq, Cheng:2017bzz}. It is thus important to see whether the LCSRs under different choices of correlator are consistent with each other.

The remaining parts of this paper are organized as follows. In Sec.~\ref{Section:2}, we present the calculation technologies for deriving the key components, e.g. the TFFs $f^{0,+}$, of the $B\to \pi$ semileptonic decays. In Sec.~\ref{Section:3}, we present our numerical results and discussions. Section.~\ref{Section:4} is reserved for a summary.

\section{Calculation technology}\label{Section:2}

For the $B\to\pi$ decays, its transition matrix element can be written as
\begin{eqnarray}
&&\langle\pi(p)|\bar{u} \gamma_\mu b |B(p+q)\rangle\nonumber\\
&=& f^+(q^2)\Big[2p_\mu + \left(1-\frac{m_B^2-m_\pi^2}{q^2}\right) q_\mu\Big] \nonumber\\
&& + f^0(q^2)\frac{m_B^2-m_\pi^2}{q^2}q_\mu,  \label{Eq:matrix1}
\end{eqnarray}
where $p$ is the momentum of pion and $p+q$ is the momentum of $B$ meson, $f^{+,0}(q^2)$ are the two TFFs. Within the HQEFT framework, the $B\to \pi$ transition matrix element can be expanded as $1/m_b$-power series and at the leading-order accuracy, we have~\cite{Wang:2001mi,Wang:2002zba}:
\begin{eqnarray}
\langle\pi(p)| \bar{u} \gamma_\mu b |B(p+q)\rangle
&=& \frac{\sqrt{m_B}}{\sqrt{\bar\Lambda_B}}\langle \pi(p)|\bar u \gamma _\mu b_v^+ |B_v \rangle \\
&=& -  \frac{\sqrt{m_B}}{\sqrt{\bar\Lambda_B}}{\rm Tr}[\pi(v,p){\gamma _\mu }B_v], \label{Eq:Bpi_HQEFT}
\end{eqnarray}
where  $\bar \Lambda_B = m_B - m_b$, $ b_v^+ $ is the effective $b$-quark field, $\pi (v,p) = {\gamma ^5}[A(v \cdot p)+ \not\!\!\hat{p} B(v \cdot p)]$, and the effective $B$-meson spin function $B_v = -\sqrt {\bar \Lambda }(1 + \not\!\! v)\gamma^5/2$ with $v$ is the $B$-meson velocity, $\hat{p}^{\mu}=p^{\mu}/(v\cdot p)$, $\bar \Lambda$ is the binding energy, and $\bar \Lambda  = \mathop {\lim }\limits_{m_b \to \infty } {\bar \Lambda _B}$. $A(v \cdot p)$ and $B(v \cdot p)$ are scale-dependent coefficient functions characterized by the heavy-to-light transition matrix elements, where $v \cdot p = (m^2_B + m^2_\pi - q^2)/(2 m_B)$. Using those equations, we obtain the HQEFT $B\to \pi\ell\bar\nu_\ell$ TFFs $f^\pm(q^2)$, i.e.
\begin{equation}
f^\pm(q^2) = \frac{\sqrt {\bar\Lambda}}{\sqrt{m_B\bar\Lambda_B}}\left[A(y)\pm \frac{m_B}{y}B(y)\right]+ \cdots, \label{Eq:fpm}
\end{equation}
where $y=v \cdot p$, the symbol $\cdots$ denotes the high-order ${\cal O}(1/{m_b})$ contributions, and the TFF $f^0(q^2)$ can be derived via the relation:
\begin{equation}
f^0(q^2) = f^+(q^2)+ \frac{q^2}{m_B^2-m_\pi^2} f^-(q^2).  \label{Eq:f0}
\end{equation}

The two coefficient functions $A(y)$ and $B(y)$ in the TFFs $f^\pm(q^2)$ can be derived by using the LCSR approach. For the purpose, we need to design a correlator, which can be generally written as
\begin{eqnarray}
&& F_\mu (p,q) = i\int d^4x e^{i(q-m_b v) \cdot x}\langle \pi (p)|T\{ j_\mu(x),j^\dag_B(0)\} |0\rangle, \nonumber\\ \label{Eq:correlator}
\end{eqnarray}
where $j_\mu(x)$ and $j_B^{\dag}(0)$ are usually chosen as $j^{\cal U}_\mu(x)=\bar u(x)\gamma_\mu b^+_v(x)$ and $j^{ \dag\cal U}_B(0)=im_b\bar b(0)^+_v\gamma_{5}q(0)$~\cite{Khodjamirian:2011ub}. It has been observed that by using $j^{\cal R}_\mu(x) = \bar u(x)\gamma _\mu (1 + \gamma_5)b^+_v(x)$ and $j_B^{\dag\cal R }(0)= m_b \bar b^+_v(0) i (1 + \gamma _5)d(0)$, one can achieve more accurate LCSRs for the TFFs with less uncertain high-twist terms~\cite{Huang:2001xb}. Even though by using the chiral currents, one may receive additional uncertainties from the scalar ($0^+$) $B$ meson resonances, which however can be absorbed into the corresponding hadronic dispersion integral via a proper choice of continuum threshold~\cite{Li:2012gr, Li:2015cta}.

There are other choices of correlator. In the paper, we adopt the above two choices of correlators to show whether the LCSRs under different correlators are consistent with each other. For convenience, we label the LCSR for $j^{\cal U}_\mu(x)$ and $j^{\dag\cal U}_B(0)$ as LCSR-${\cal U}$, and the LCSR for $j^{\cal R}_\mu(x)$ and $j^{\dag\cal R}_B(0)$ as LCSR-${\cal R}$.

The calculation technology for deriving the LCSR-${\cal R}$ and LCSR-${\cal U}$ are the same. As an explicit example, we describe the procedures for deriving LCSR-${\cal R}$ of the two coefficient functions $A(y)$ and $B(y)$ by using the correlator $F_\mu^{\cal R}(p,q)$ with two right-handed chiral currents $j^{\cal R}_\mu(x)$ and $j^{\dag\cal R}_B(0)$.

First, we discuss the hadronic representation for the correlator. One can insert a complete series of the intermediate hadronic states in the correlator and isolate the pole term of the lowest pseudoscalar state to get the hadronic representation. That is, the correlator $F_\mu^{\cal R}$ can be expressed as
\begin{eqnarray}
&& F_\mu^{\cal R} (p,q)= \frac{ \langle \pi(p)|\bar u \gamma_\mu b^+_v |B_v \rangle \langle B_v | \bar b^+_v i \gamma_5 d | 0 \rangle }{m_B^2 - (p + q)^2} \cr
&&\quad + \sum \limits_{H_v} \frac{\langle \pi (p)|\bar u \gamma _\mu (1 + \gamma_5) b^+_v | H_v \rangle \langle H_v|\bar b^+_v i (1 + \gamma_5) d|0\rangle }{m_{H_v}^2 - (p + q)^2}.\nonumber\\
\end{eqnarray}
Based on the original definition, the matrix element $\langle B_v |\bar b^+_v i \gamma_5 d | 0 \rangle  = F {\rm Tr}[\gamma_5 B_v]/2$, where $F$ represents the leading-order $B$-meson decay constant, which can be found in Refs.~\cite{Wang:2001mi, Wang:2002zba}. With the help of the $B\to\pi$ matrix element \eqref{Eq:Bpi_HQEFT}, the correlator $F_\mu^{\cal R}$ can be written as
\begin{eqnarray}
F_\mu^{\cal R} (p,q)  &=& 2 i F \frac{A(y)v^\mu  + B(y) \hat p ^\mu }{2 \bar\Lambda_B - 2v \cdot k} \cr
&& + \int_{s_0}^\infty  ds \frac{\rho (y,s)}{s - 2v \cdot k} + {\rm Subtractions}, \label{Eq:Fpq_HQEFT}
\end{eqnarray}
where $k^\mu = P^\mu_B - m_b v^\mu$ is the heavy hadronic residual momentum. The spectral density $\rho (y,s)$ can be estimated by using the ansatz of the quark-hadron duality~\cite{Shifman:1978bx, Shifman:1978by}. Contributions from the higher power suppressed subtraction terms can be suppressed or eliminated by applying the Borel transformation.

On the other hand, we need to deal with the corrector in large spacelike momentum regions $(p+q)^2-m_b^2 \ll 0$ and $q^2 \ll m_b^2$ for the momentum transfer, which correspond to the small light-cone distance $x^2 \simeq 0$ and are required by the validity of operator product expansion (OPE). Based on the OPE, the correlator can be expanded into a power series of the pion LCDAs with various twists:
\begin{eqnarray}
F_\mu^{\cal R} (p,q) &=& i\int d^4x e^{iq \cdot x}\int_0^\infty  dt \delta (x - vt)\nonumber\\
&\times&\langle \pi (p)|T \bar u(x) \gamma _\mu \gamma _5 d(0) |0\rangle,
\end{eqnarray}
where we have implicitly used the $B$-meson heavy-quark propagator within HQEFT, i.e. $S(x,v)=(1+\slash\!\!\! v )\times\int_0^\infty dt \delta(x-vt)/2$. For convenience, we present the needed pion LCDAs up to twist-4 accuracy in the Appendix, which are taken from Refs.\cite{Braun:1989iv, Ball:1998je}.

As a final step, by applying the Borel transformation for the correlator $F^{\cal R}_\mu (p,q)$, we obtain the LCSRs for the coefficient functions $A(y)$ and $B(y)$,
\begin{eqnarray}
A(y) &=& -\frac{f_\pi}{2F}\int_0^{s_0} ds e^{(2 \bar \Lambda _B - s)/T} \frac{1}{y^2} \frac{\partial}{\partial u} g_2(u) \bigg|_{u = 1 - \frac{s}{2y}},
\label{Eq:Ay}\\
B(y) &=& -\frac{f_\pi}{2F}\int_0^{s_0} ds e^{(2 \bar \Lambda _B - s)/T} \bigg[- \phi_{2;\pi}(u)
\nonumber\\
&& + \bigg(\frac{1}{y}\frac{\partial }{\partial u}\bigg)^2 g_1(u) - \frac{1}{y^2}\frac{\partial}{\partial u}g_2(u)\bigg]_{u = 1 - \frac{s}{2y}}.\label{Eq:By}
\end{eqnarray}
Substituting them into Eqs.(\ref{Eq:fpm}, \ref{Eq:f0}), one can obtain the expressions for the needed two TFFs $f^+(q^2)$ and $f^0(q^2)$, i.e.
\begin{eqnarray}
f^+(q^2) &=&  - \frac{f_\pi\sqrt{\bar\Lambda}}{2F\sqrt{m_B\bar\Lambda_B}}  \int_0^{s_0} ds e^{(2 \bar \Lambda_B - s)/T} \nonumber\\
&&\times\bigg\{\frac{1}{y^2}\frac{\partial }{\partial u}g_2(u) + \frac{m_B}{y} \bigg[ - \phi_{2;\pi}(u) \nonumber\\
&&+ \bigg(\frac{1}{y} \frac{\partial}{\partial u}\bigg)^2 g_1 (u) - \frac{1}{y^2}\frac{\partial }{\partial u}g_2(u)\bigg]\bigg\}\bigg|_{u = 1 - \frac{s}{2y}}, \nonumber\\
f^0(q^2) &=&  - \frac{f_\pi\sqrt{\bar\Lambda}}{2F\sqrt{m_B\bar\Lambda_B}}  \int_0^{s_0} ds e^{(2 \bar \Lambda_B - s)/T} \nonumber\\
&&\times\bigg\{\bigg(1 + \frac{q^2}{m_B^2 - m_{\pi}^2}\bigg)\frac{1}{y^2}\frac{\partial }{\partial u}g_2(u) \nonumber\\
&&+ \bigg(1 - \frac{q^2}{m_B^2-m_{\pi}^2}\bigg)\frac{m_B}{y} \bigg[ - \phi_{2;\pi}(u) \nonumber\\
&&+ \bigg(\frac{1}{y} \frac{\partial}{\partial u}\bigg)^2 g_1 (u) - \frac{1}{y^2}\frac{\partial }{\partial u}g_2(u)\bigg]\bigg\}\bigg|_{u = 1 - \frac{s}{2y}}.
\end{eqnarray}

Following the same calculation technology, the LCSRs for the usual correlator $F_\mu^{\cal U}(p,q)$ can also be obtained, e.g.
\begin{eqnarray}
f^+(q^2) &=&  - \frac{f_\pi\sqrt{\bar\Lambda}}{4F\sqrt{m_B\bar\Lambda_B}}  \int_0^{s_0} ds e^{(2 \bar \Lambda_B - s)/T} \nonumber\\
&&\times\bigg\{\frac{1}{y^2}\frac{\partial }{\partial u}g_2(u)-\frac{\mu_\pi}{y} \phi_p(u) -\frac{\mu_\pi}{6y} \frac{\partial}{\partial u}\phi_\sigma(u)\nonumber\\
&&+ \frac{m_B}{y} \bigg[ - \phi_{2;\pi}(u) + \bigg(\frac{1}{y} \frac{\partial}{\partial u}\bigg)^2 g_1 (u) \nonumber\\
&&- \frac{1}{y^2}\frac{\partial }{\partial u}g_2(u)+\frac{\mu_\pi}{6y} \frac{\partial}{\partial u}\phi_\sigma(u)\bigg]\bigg\}\bigg|_{u = 1 - \frac{s}{2y}}, \nonumber\\
f^0(q^2) &=&  - \frac{f_\pi\sqrt{\bar\Lambda}}{4F\sqrt{m_B\bar\Lambda_B}}  \int_0^{s_0} ds e^{(2 \bar \Lambda_B - s)/T} \nonumber\\
&&\times\bigg\{\bigg(1 + \frac{q^2}{m_B^2 - m_{\pi}^2}\bigg) \nonumber\\
&&\times\bigg(\frac{1}{y^2}\frac{\partial }{\partial u}g_2(u)-\frac{\mu_\pi}{y} \phi_p(u) -\frac{\mu_\pi}{6y} \frac{\partial}{\partial u}\phi_\sigma(u) \bigg) \nonumber\\
&&+ \bigg(1 - \frac{q^2}{m_B^2-m_{\pi}^2}\bigg)\frac{m_B}{y} \bigg[ - \phi_{2;\pi}(u) \nonumber\\
&&+ \bigg(\frac{1}{y} \frac{\partial}{\partial u}\bigg)^2 g_1 (u) - \frac{1}{y^2}\frac{\partial }{\partial u}g_2(u)\nonumber\\
&&+\frac{\mu_\pi}{6y} \frac{\partial}{\partial u}\phi_\sigma(u)\bigg]\bigg\}\bigg|_{u = 1 - \frac{s}{2y}}.
\end{eqnarray}

The Borel parameter $T$ and the continuum threshold $s_0$ are determined such that the resulting form factor does not depend too much on the precise values of
these parameters; in addition the continuum contribution, which is the part of the dispersive integral from $s_0$ to $\infty$ that has been subtracted from both sides of the equation, should not be too large.

\section{Numerical analysis}\label{Section:3}

\subsection{Input parameters}

As for the input parameters, the masses of pion, $B$-meson and $b$ quark are taken as $m_{\pi^+}=0.1395$ GeV, $m_B=5.279$ GeV and $m_b=4.75\pm{0.05}$ GeV, which are from the particle data group~(PDG)~\cite{2018pdg}. As for the decay constant, we take the pion decay constant $f_{\pi}=0.1304$ GeV~\cite{2018pdg} and the leading order $B$-meson effective decay constant $F=0.34\pm{0.04}$ $\rm{GeV^{3/2}}$~\cite{Wang:2000sc}. The factorization scale $\mu$ is set as typical momentum of the $B$ meson decays, e.g. $\mu_b ={(m_B^2 - m_b^2)^{1/2}} \approx 2.4$ GeV, and we set its error as $\Delta \mu = \pm 1$ GeV.

For the twist-2 LCDA $\phi_{2;\pi}$, we adopt the Brodsky-Huang-Lepage (BHL) model~\cite{BHL_1} to do this calculation. The BHL prescription of the hadronic wave function (WF) is obtained by connecting the equal-time WF in the rest frame and the WF in the infinite momentum frame. Meanwhile this LCDA can mimic the DA behavior from asymptotic-like to CZ-like naturally. Then with the experimental data on various processes, one may decide which is the right DA behavior possessed by the light pseudo-scalar mesons. And refs.~\cite{Wu:2010zc,Wu:2011gf} suggested that considering BHL model with the Wiger-Melosh rotation effects, they can get meaningful results which coincide with the experiment constrains. Furthermore, the process of the calculation of pion twist-2 LCDA can be found in the ref.~\cite{Huang:2013yya}. After integrating the transverse momentum part out, the twist-2 LCDA takes the form
\begin{eqnarray}
\phi _{2;\pi}(u,\mu^2) & = & \frac{\sqrt 3 Am_q\beta}{2\pi ^{3/2}f_\pi}\sqrt{u \bar u} \varphi (u) \nonumber\\
&\times& \bigg\{ {\rm Erf} \left[\sqrt \frac{m_q^2 + \mu^2}{8\beta ^2 u\bar u} \right] - {\rm Erf} \left[\sqrt \frac{m_q^2}{8\beta^2 u\bar u} \right] \bigg\},\nonumber\\
\end{eqnarray}
where $\bar{u} = 1 - u$, $\xi  =  2u - 1$, $\varphi (u) = 1 + B \times C_2^{3/2}(\xi)$ and ${\rm Erf}(x) = 2\int_0^x e^{-t^2} dt/\sqrt \pi$. The light constitute quark mass $m_{q}\simeq 0.30~{\rm GeV}$, and the remaining three parameters $A$, $\beta$ and $B$ can be fixed by using three constraints~\cite{Huang:2013yya, BHL_1}:
\begin{itemize}
\item[(i)] The pion wave function normalization condition,
\begin{equation}
\int_0^1 {dx} \int {\frac{d^2{\bf k}_\bot}{16\pi ^3}\Psi _\pi} (x,{\bf k}_ \bot) = \frac{f_\pi }{2\sqrt 6 };
\end{equation}
\item[(ii)] The sum rule derived from $\pi^{0} \to \gamma \gamma$ decay amplitude implies,
\begin{equation}
\int_0^1 dx \Psi _\pi(x,{\bf k}_\bot = {\bf 0}) = \frac{\sqrt 6}{f_\pi};
\end{equation}
\item[(iii)] Generally, the twist-2 LCDA can be expanded as a Gegenbauer polynomial. We can get the second moment $a_2$ by using the orthogonality relation,
\begin{equation}
a_n(\mu ^2) = \frac{\int_0^1 du \phi _\pi(u,\mu ^2)C_n^{3/2}(2u-1)}{\int_0^1 du 6u\bar u[C_n^{3/2}(2u-1)]^2}.
\end{equation}
\end{itemize}

\begin{table}[htb]
\begin{center}
\begin{tabular}{c c c c  }
\hline
~~$a_2(\mu^2_0)$~~&~~$A$~~&~~$\beta$~~&~~$B$~~\\
\hline
~~0.039~~&~~23.50~~&~~0.61~~&~~0.075~~\\
~~0.112~~&~~24.63~~&~~0.59~~&~~0.010~~\\
~~0.185~~&~~23.50~~&~~0.63~~&~~0.141~~\\
\hline
\end{tabular}
\caption{Parameters of the twist-2 LCDA $\phi _{2;\pi}(u,\mu_0^2)$ determined for $a_2(\mu^2_0)=0.112\pm{0.073}$~\cite{Huang:2013yya}.}\label{para}
\end{center}
\end{table}

\begin{figure}[htb]
\begin{center}
\includegraphics[width=0.45\textwidth]{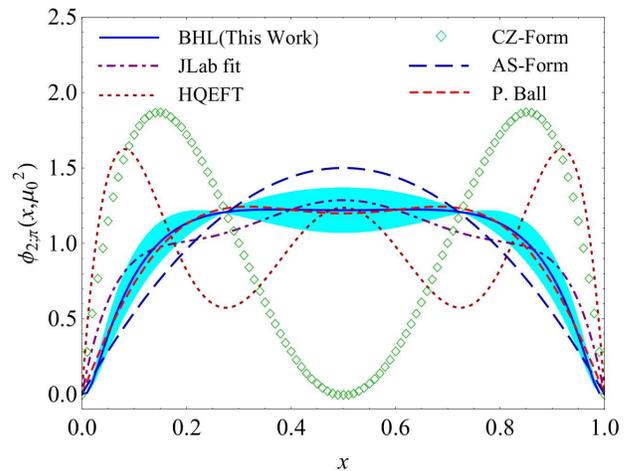}
\end{center}
\caption{The behavior of LCDA $\phi_{2;\pi}(x,\mu^2_0)$ predicted by BHL model, in which the uncertainties are coming from all mentioned input parameters. As a comparison, we also present the JLab fit~\cite{Khodjamirian:2011ub}, QCD SR~\cite{Ball:2004ye}, CZ form~\cite{Chernyak:1981zz} and the Asymptotic form~\cite{Lepage:1980fj}.} \label{DA}
\end{figure}

The Gegenbauer moment $a_2$ can be fixed via a global fit of various experimental data involving pion, and we adopt the result given in Ref.\cite{Huang:2013yya}, e.g. $a_2(\mu^2=1\rm{GeV^2})=0.112\pm{0.073}$. The determined parameters at the scale of $1\rm{GeV}$ are presented in Table.~\ref{para}, and one can get their values at any other scale by using the renormalization group evolution. At present, there is no definite conclusion on whether the pion LCDA $\phi_{2;\pi}(u,\mu_0^2)$ is CZ-form~\cite{Chernyak:1981zz}, asymptotic form~\cite{Lepage:1980fj} or even flat-like~\cite{RuizArriola:2002bp}. We present the twist-2 DAs in Fig.~\ref{DA}, where the asymptotic form, the CZ-form, the QCD SVZ sum rule prediction~\cite{Ball:2004ye} and the prediction by A. Khodjamirian {\it et al.}~\cite{Khodjamirian:2011ub} via using the Jefferson Lab data on $F_\pi$ are presented as a comparison. By varying $a_2$, the $\phi_{2;\pi}$ shape shall varies from a single peaked behavior to double humped behavior; The present adopted BHL-model provides a convenient way to mimic the pion twist-2 LCDA behavior and we shall adopt it together with error to discuss the theoretical uncertainty caused by the different choice of $\phi_{2;\pi}$.

\subsection{The $B \to \pi$ TFFs}

As for the $B \to \pi$ TFFs $f^{+/0}(q^2)$, the allowable range of the Borel parameter $T$, and the effective continuous threshold $s_0$ are determined from the following three conditions:
\begin{itemize}
\item The continuum contribution is not higher than $35\%$ of the total LCSR;
\item All high-twist LCDAs contribution does not exceed $50\%$ of the total LCSR;
\item The derivatives of LCSRs for TFFs with respect to $(-1/T)$ give LCSR for the $B$-meson mass $m_B$; we require the estimated $B$-meson mass to be fulfilled in comparing with the experiment one, e.g. $|m^{\rm th}_B-m^{\rm exp}_B|/m^{\rm exp}_B$ less than $0.1\%$.
\end{itemize}
The flatness of the TFFs versus the Borel parameter $T$ provides an extra weaker constraint for the range of $T$. With these criteria for the LCSRs, the determined continuum threshold $s_0$ and the Borel parameter $T$ for the $B \to \pi$ vector TFF $f^+(q^2)$ are $s_0 = 2.0\pm{0.1}\;{\rm GeV}^2$ and $T = 1.4\pm{0.1}\; {\rm GeV}^{2}$ for the LCSR-$\mathcal{R}$; and $s_0 = 2.2\pm{0.1}\;{\rm GeV}^2$ and $T = 1.5\pm{0.1}\;{\rm GeV}^2$ for the LCSR-$\mathcal{U}$.

\begin{table}[htb]
\begin{center}
\begin{tabular}{c c c c c c c}
\hline
 &~~Total~~&~$\Delta \mu$~&~$\Delta$ DA~&~$\Delta T$~&~$\Delta s_0$~&~$\Delta(m_b;F)$~~\\
\hline
$f_{\mathcal{R}}^+(0)$~&~$0.276^{+0.022}_{-0.026}$~&~$^{+0.000}_{-0.001}$~&~$^{+0.017}_{-0.016}$~&~$^{+0.001}_{-0.001}$~&~$^{+0.011}_{-0.011}$~&~$^{+0.015}_{-0.011}$ \\
$f_{\mathcal{U}}^+(0)$~&~$0.282^{+0.016}_{-0.020}$~&~$^{+0.000}_{-0.000}$~&~$^{+0.009}_{-0.008}$~&~$^{+0.002}_{-0.001}$~&~$^{+0.007}_{-0.007}$~&~$^{+0.017}_{-0.012}$ \\
\hline
\end{tabular}
\caption{Uncertainties of the TFFs $f_{\mathcal{R/U}}^+(q^2)$ at $q^2 = 0$. The uncertainties of total TFFs from the mentioned input parameters are summed up in quadrature.}
\label{uncertainties}
\end{center}
\end{table}

In Table.~\ref{uncertainties}, one show how the errors of the TFFs cased by the uncertainties of the parameters, the factorization scale $\mu$, the Borel parameter $T$, the continuum threshold $s_0$, the $b$-quark mass $m_b$, the $B$-meson effective decay constant $F$, and the twist-2 LCDA. Table.~\ref{uncertainties} shows that the main errors come from the choices of $m_b$, $F$ and LCDA, whose effect could be up to $\sim12\%$. Because $m_b$ and $F$ are correlated with each other, e.g. a larger $m_b$ could result in a smaller $F$~\cite{Wu:2009kq}, we change their value simultaneously to discuss their effect to the TFFs.

\begin{table}[htb]
\begin{center}
\begin{tabular}{c c c c c}
\hline
 &~~Twist-2~~&~~Twist-3~~&~~Twist-4~~&~~Central~~\\
\hline
$f_{\mathcal{R}}^+(0)$&$0.284$&~-~&$-0.009$&0.276   \\
$f_{\mathcal{U}}^+(0)$&$0.153$&$0.134$&$-0.004$&0.282   \\
\hline
\end{tabular}
\caption{Different twist terms to the $B \to \pi$ TFFs $f_{\mathcal{R/U}}^+$ at the large recoil region $q^2 \simeq 0 ~\rm{GeV^2}$.}
\label{con}
\end{center}
\end{table}

We present the contributions of different twist terms to the dominant $B \to \pi$ TFFs $f_{\mathcal{R/U}}^+$ at the large recoil region $q^2 \simeq 0 ~\rm{GeV^2}$ in Table.~\ref{con}. For the case of usual correlator $F^{\cal U}_\mu$, contributions of the twist-3 terms are about $45\%$ of the TFF $f_{\mathcal{U}}^+(0)$. The twist-3 DAs are not known as well as the DAs of twist-2.  And for the case of U correlator, Table III shows that the twist-2 and twist-3 terms have same size; thus, by using the R correlator, one can achieve a much better accuracy LCSR for the transition form factor. Table II shows the errors for the case of U correlator are seemingly small, this is because we do not include the uncertainty from twist-3 DAs in the present prediction. And twist-3 DAs uncertainties is not the point which this study focuses on. Meanwhile by using the chiral current correlator, we could get the precision twist-2 DAs information, in turn with the experiment data the twist-2 DAs would be determined exactly. And it is hard to distinguish the information of twist-2 or twist-3 in the usual current, because their contribution are the same size and blend with each other. While for the case of the right-handed chiral correlator $F^{\cal R}_\mu$, contributions from the twist-3 terms are eliminated. Moreover, the twist-4 terms are negligible for both cases. Thus by taking the chiral correlator, uncertainties from the higher-twist LCDAs are greatly suppressed, leading to a more precise LCSR prediction. By summing the contributions from various twist terms, the total TFFs for both correlators agree well with each other with errors \footnote{The differences between their central values, whose magnitude is comparable to the magnitude of the twist-4 terms, could be treated as the difference of the uncalculated high-twist terms.}. The LCSR approximant should not depend on the choice of chiral correlator, and our present prediction for the TFFs provide one of such examples.

The LCSR prediction is valid when the final-state pion has large energy in the rest-system of $B$-meson, e.g. $E_{\pi} \gg \Lambda_{\rm{QCD}}$. Using the relation, $q^2=m^2_B-2m_B E_{\pi}$, we adopt a conservative range, $0< q^2 < 12\; \rm{GeV^2}$, as the applicable range of the LCSR method. Thus before dealing with the $B\to\pi$ semi-leptonic decay, we need to extend the TFFs to the physically allowable range. We use the $z$-series parametrization to extrapolate the calculated TFFs to the physically allowable $q^{2}$-region~\cite{Khodjamirian:2011ub}, e.g.
\begin{eqnarray}
f^+ (q^2) &=& \frac{f^+ (0)}{1 - q^2/m_{B^*}^2}\bigg\{ 1 + \sum_{k = 1}^{N - 1} b_k \bigg[z(q^2)^k - z(0)^k\nonumber\\
 &&- ( - 1)^{N - k}\frac{k}{N}(z(q^2)^N - z(0)^N)\bigg]\bigg\}
\end{eqnarray}
and
\begin{equation}
f^0(q^2) = f^0(0)\bigg\{ 1 + \sum_{k = 1}^{N - 1} b_k (z(q^2)^k - z(0)^k)\bigg\}.
\end{equation}
The function $z(q^2)$ is defined as
\begin{equation}
z(q^2) = \frac{\sqrt {t_ +  - q^2}  - \sqrt {t_ +  - t_0} }{\sqrt {t_ + - q^2}  + \sqrt {t_ + - t_0} },
\end{equation}
where $t_ \pm = (m_B \pm m_\pi)^2$, $t_0 = t_ + (1 - \sqrt {1-t_+/t_-} )$. The coefficients $b_0=f^+(0)$, $b_1$, $b_2$ and $b_3$ are determined such that the quality of fit $(\Delta)$ is less than $1\%$. The quality of fit is defined as:
\begin{equation}
\Delta  = \frac{\sum_t {|f^{i} - f^{\rm fit}|}}{\sum_t {|f^{i}|} } \times 100\%,t \in \bigg\{ 0,\frac{1}{2},...,\frac{23}{2},12\bigg\} ~\rm{GeV}^2.\nonumber
\end{equation}

\begin{table}[!htb]
\begin{center}
\begin{tabular}{c c c c c c }
\hline
~~~&~~~$b_1$~~~&~~~$b_2$~~~&~~~$b_3$~~~&~~~$\Delta$~~~~~\\
\hline
$f_{\mathcal{R}}^+(q^2)$& $-1.600$ & $-1.453$ &~-~& $0.03\%$ \\
$f_{\mathcal{R}}^0(q^2)$& $-1.155$ & $2.075$ & $-3.377$ & $0.01\%$ \\
$f_{\mathcal{U}}^+(q^2)$& $-1.309$ & $-1.757$ &~-~& $0.50\%$\\
$f_{\mathcal{U}}^0(q^2)$& $-1.197$ & $0.656$ & $-0.611$ & $0.06\%$ \\
\hline
\end{tabular}
\caption{Fitted parameters and qualities of fit of LCSR-$\mathcal{R}$ and LCSR-$\mathcal{U}$ TFFs $f^{+/0}(q^2)$.}\label{fitted}
\end{center}
\end{table}

The determined parameters for the $z$-series and the quality of fit are listed in Table.~\ref{fitted}.

\begin{figure*}[htb]
\begin{center}
\includegraphics[width=0.45\textwidth]{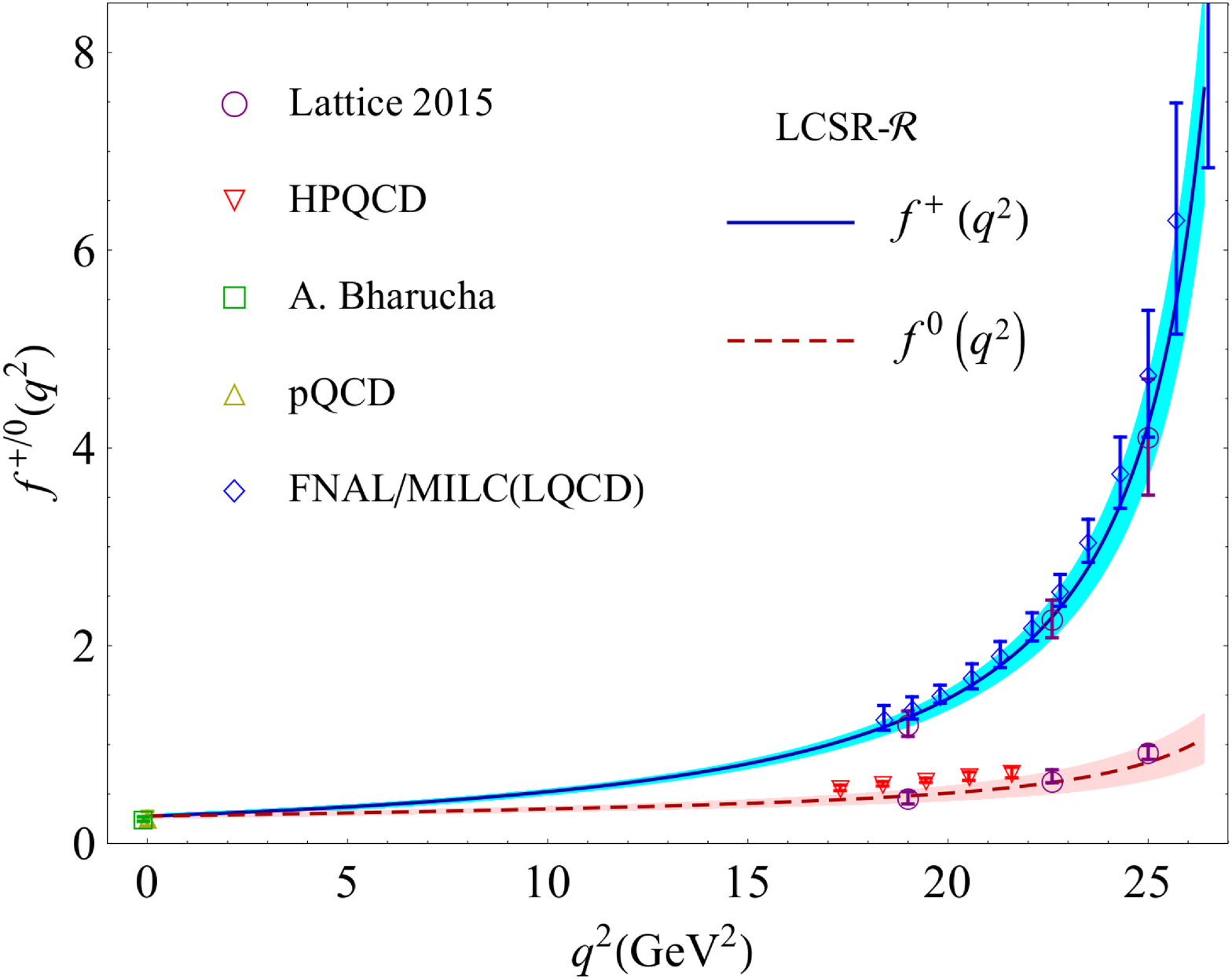} ~~~~~\includegraphics[width=0.45\textwidth]{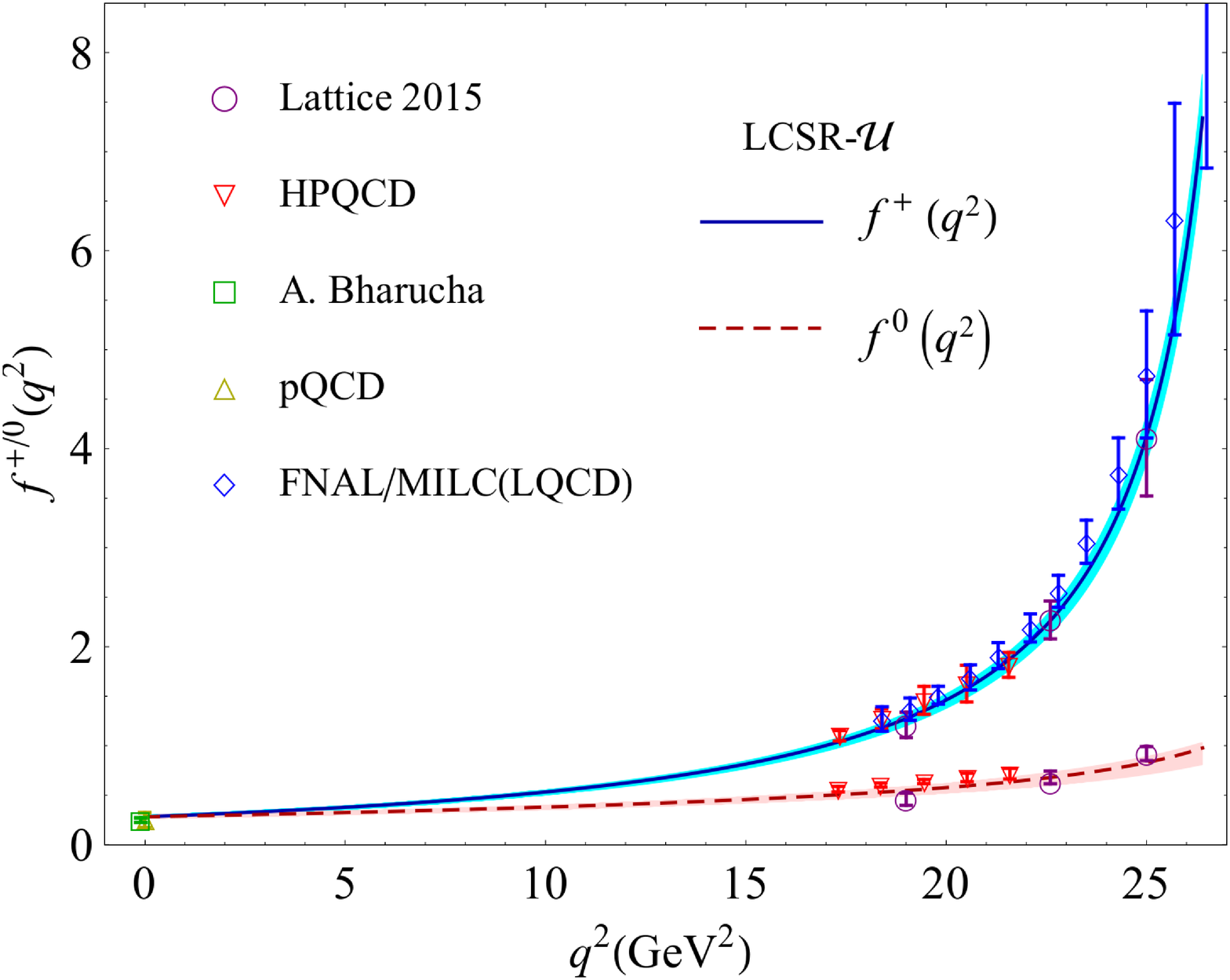}
\end{center}
\caption{The extrapolated $B\to\pi$ TFFs $f^{+/0}(q^2)$ calculated under two different correlators, where the left/right one stands for the LCSR-${\cal R}$/${\cal U}$, respectively. The shaded bands are uncertainties from the mentioned error sources. As a comparison, we also present the predictions coming from Lattice QCD~\cite{Flynn:2015mha}, LCSR~\cite{Bharucha:2012wy,Imsong:2014oqa}, pQCD~\cite{Li:2012nk} and HPQCD~\cite{Dalgic:2006dt}. }
\label{f+}.
\end{figure*}

The extrapolated HQEFT LCSRs for the $B \to \pi$ TFFs $f^{+/0}(q^2)$ are presented in Fig.(\ref{f+}), where left/right diagram stands for the LCSR-${\cal R}$/${\cal U}$, respectively. In Fig.(\ref{f+}), the solid lines are TFFs with all parameters to be their central values, and the shade bands are TFF errors that are squared averages of the errors caused by all the error sources. Our predictions within errors for $f^{+/0}(q^2)$ are consistent with the Lattice QCD predictions obtained by the HPQCD and Fermilab/MILC collaborations.

We adopt the correlation coefficient $\rho_{XY}$ to measure to what degree the TFFs under different choices of chiral correlators are correlated with each other, which is defined as~\cite{2018pdg},
\begin{equation}
\rho_{XY} = \frac{{\rm Cov}(X,Y)}{\sigma_{X}\sigma_{Y}},
\end{equation}
where $X$ and $Y$ stand for the corresponding LCSR-$\mathcal{R}$ and LCSR-$\mathcal{U}$ TFFs, respectively. The covariance is $ {\rm Cov}(X,Y)=E[( X-E(X)(Y-E(Y))]=E(XY)-E(X)E(Y)$, where $E$ is the expectation value of $X$ or $Y$. $\sigma_{X} $ and $\sigma_{Y} $ are standard deviations of $X$ and $Y$. The rang of correlation coefficient $|\rho_{XY}|$ is 0 $\sim$ 1. A larger $|\rho_{XY}|$ indicates a higher consistency between $X$ and $Y$. The correlation coefficient $|\rho_{RU}|$ of LCSR-$\mathcal{R}$ and LCSR-$\mathcal{U}$ is 0.85 for the vector TFF $f^+(q^2)$, which changes to 0.52 for the scalar TFF $f^0(q^2)$. The correlation coefficient of both TFFs are lager than 0.5, implying those TFFs are consistent with each other.   \\

\subsection{The $B \to \pi \ell\bar \nu_\ell $ differential branching fraction, the matrix element $|V_{ub}|$  and the ratio $\mathcal{R}_{\pi}$}

\begin{figure*}[htb]
\begin{center}
\includegraphics[width=0.45\textwidth]{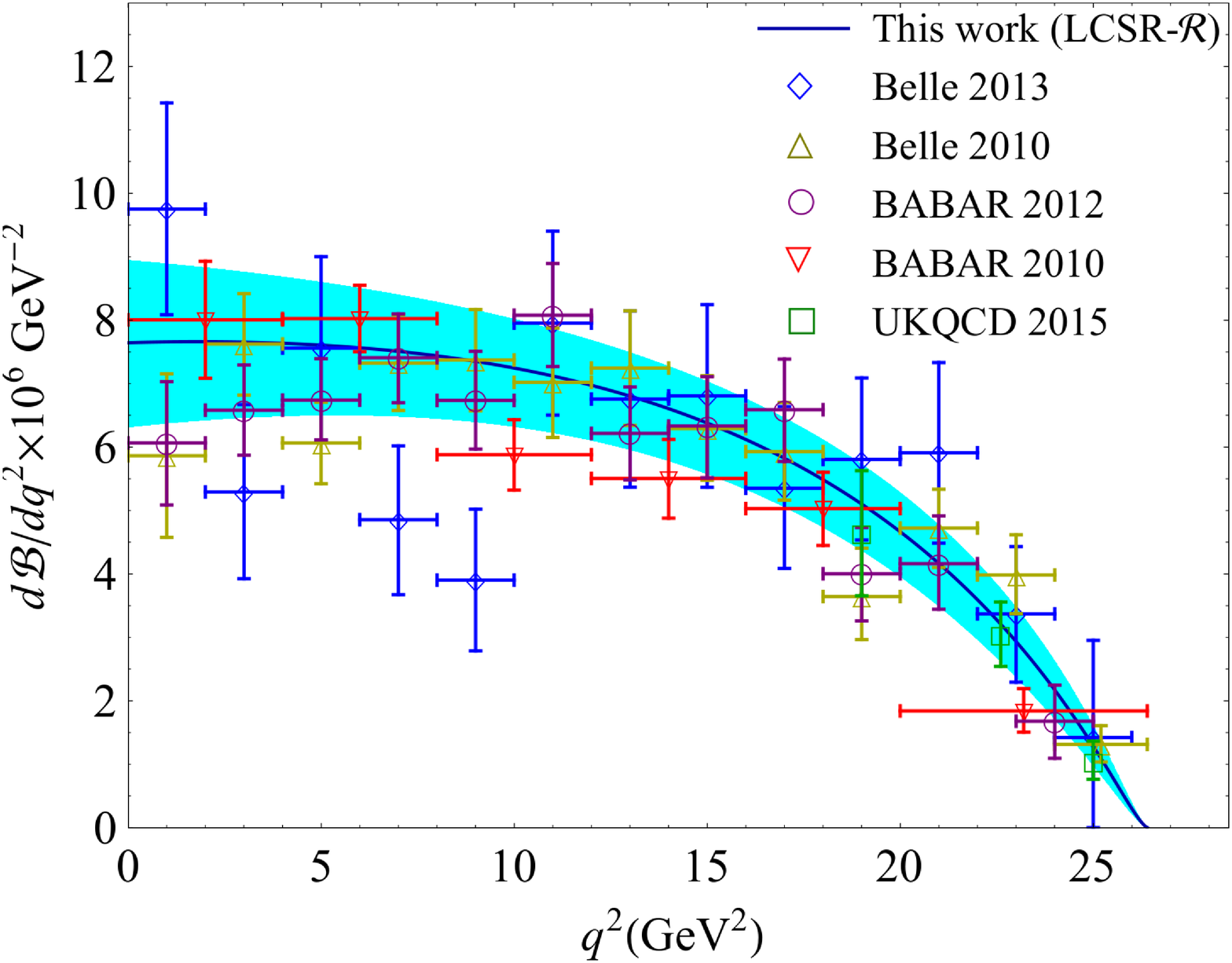} ~~~~~\includegraphics[width=0.45\textwidth]{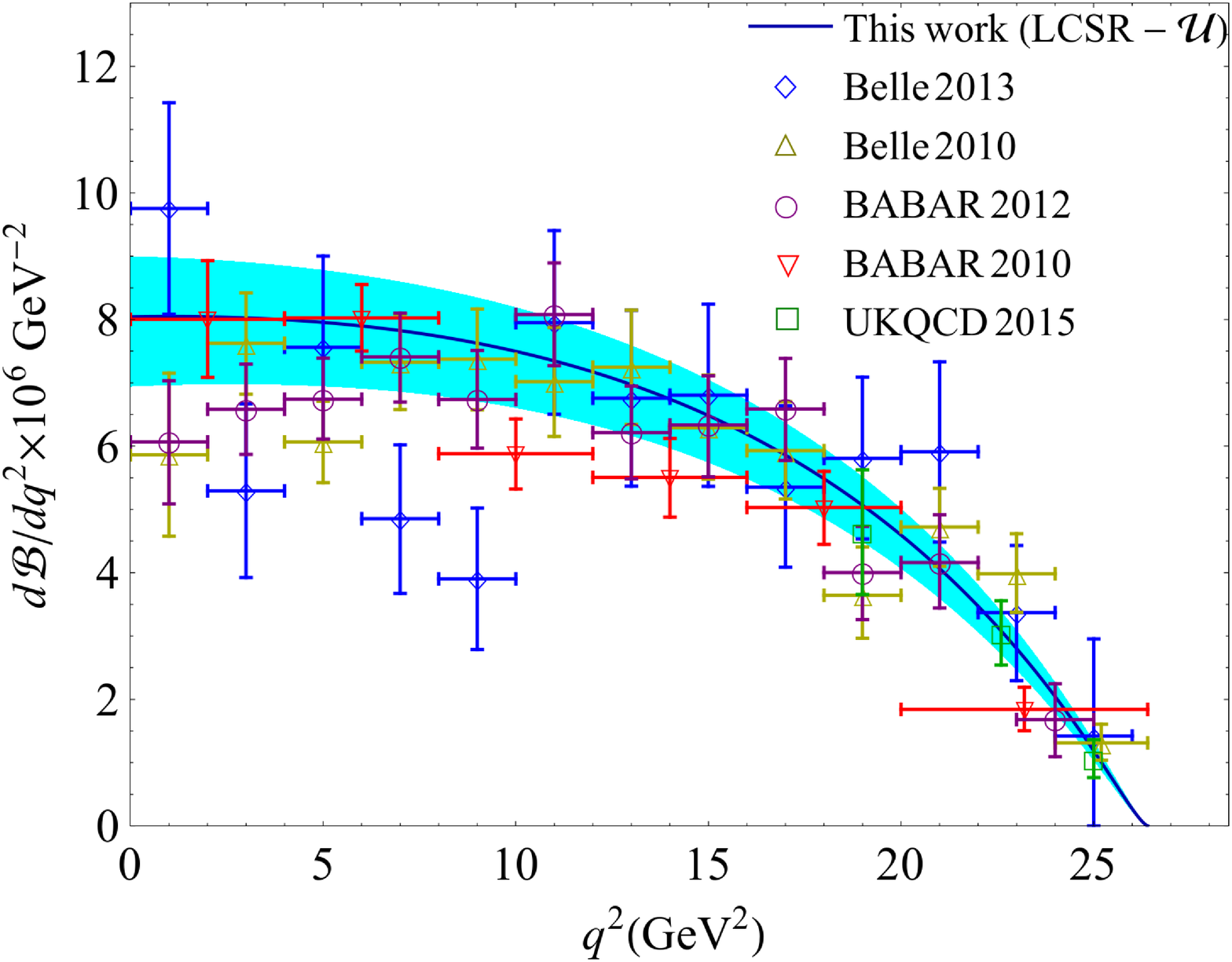}
\end{center}
\caption{The $B \to \pi \ell \bar\nu_\ell$ differential branching fractions for two different correlators, in which the left/right one stands for LCSR-${\cal R}$/${\cal U}$, respectively. As a comparison, we also present the BABAR data~\cite{Lees:2012vv, delAmoSanchez:2010af}, and the Belle data~\cite{Ha:2010rf, Sibidanov:2013rkk}. } \label{dbdq}
\end{figure*}

The $B \to \pi \ell\bar \nu_\ell $ differential decay widths can be written as~\cite{Li:2008tk,Colangelo:2006vm},
\begin{widetext}
\begin{eqnarray}
\frac{d\Gamma(B \to \pi \ell\bar \nu_\ell )}{dq^2}&=&\frac{G_F^2|V_{ub}|^2}{192 \pi^3 m_B^3}\frac{q^2-m_\ell^2}{(q^2)^2}\sqrt{\frac{\left (q^2-m_\ell^2 \right )^2}{q^2}} \times \sqrt{\frac{\left (m_B^2-m_\pi^2-q^2 \right )^2}{4q^2}-m_\pi^2}\Bigl \{ \left (m_\ell^2+2q^2 \right )\nonumber\\
&\times& \left [q^2-\left (m_B-m_\pi \right )^2 \right ]\left [ q^2-\left (m_B+m_\pi \right )^2 \right ] \times f_+^2(q^2) + 3m_\ell^2\left (m_B^2-m_\pi^2 \right )^2 f_0^2(q^2)\Bigr \},
\end{eqnarray}
\end{widetext}
where $m_l$ is the lepton mass. For the light lepton $l = e $ or $\mu$, the lepton mass can be safely neglected, and the differential decay width changes to~\cite{Li:2008tk, Colangelo:2006vm}
\begin{equation}
\frac{d\Gamma(B \to \pi \ell\bar \nu_\ell )}{dq^2}= \frac{G_F^2
  |V_{ub}|^2}{ 192 \pi^3 m_B^3}\,\lambda^{3/2}(q^2) |f_+(q^2)|^2 ,
\end{equation}
where the $|V_{ub}|$ is the CKM matrix element, $G_F$ is the Fermi-coupling constant, and the phase-space factor $\lambda(q^2) = (m_B^2+m_\pi^2-q^2)^2 - 4 m_B^2 m_\pi^2$. The branching fraction has the relation with the decay width, $d{\cal B}/dq^2 = d\Gamma(B \to \pi \ell\bar \nu_\ell)/dq^2\times \tau_{B^0} $, where the total lifetime $\tau_{B^0}=1.525\pm{0.009}$ ps~\cite{2018pdg}. Fig.(\ref{dbdq}) indicates that both the LCSR-${\mathcal R}$ and the LCSR-${\mathcal U}$ predictions together with their errors ( shown by shaded bands) are consistent with the BABAR and BELLE data.

Experimentally, people have measured the integrated branching fraction, $\Delta {\cal B}(0,q_f^2)=\int_{0}^{q_f^2} (d{\cal B}/dq^2) dq^2$~\cite{Lees:2012vv, Sibidanov:2013rkk},
\begin{eqnarray}
\Delta {\cal B}(0,12\rm{GeV}^2) &=& (0.83 \pm 0.03 \pm 0.04) \times 10^{ - 4},\nonumber\\
&&({\rm BABAR}~ 2012) \label{br1}\\
\Delta {\cal B}(0,12\rm{GeV}^2) &=& (0.808 \pm 0.06) \times 10^{ - 4}.\nonumber\\
&&({\rm Belle}~ 2013) \label{br2}
\end{eqnarray}
Using their weighted averages, we can inversely determine the value of $|V_{ub}|$, e.g.
\begin{eqnarray}
|V_{ub}|_{\cal R} &=&  (3.45^{+0.28}_{-0.20}\pm{0.13}_{\rm{exp}})\times10^{-3},\\
|V_{ub}|_{\cal U} &=& (3.38^{+0.22}_{-0.16}\pm{0.12}_{\rm{exp}})\times10^{-3},
\end{eqnarray}
where the first error is squared average of those from the input parameters and the second error is the experimental error of $\Delta {\cal B}(0,12\rm{GeV}^2)$.

\begin{table}[htb]
\begin{center}
\begin{tabular}{ c c }
\hline
~~~~~~~~~~~Exclusive decays~~~~~~~~~~~~~~ &  ~~~~~~~~~~~~~~~~~~~ \\
$B\rightarrow \pi\ell \bar{\nu}_{\ell} $ & \raisebox {2.0ex}[0pt]{$|V_{ub}|\times 10^{3}$} \\
\hline
LCSR-$\mathcal{R}$ (This work)  &  $3.45{^{+0.28}_{-0.20}\pm{0.13}_{\rm{exp}}}$     \\
LCSR-$\mathcal{U}$ (This work) &   $3.38{^{+0.22}_{-0.16}\pm{0.12}_{\rm{exp}}}$     \\
RBC/UKQCD ~\cite{Flynn:2015mha} &  $3.61\pm0.32$\\
Fermilab/MILC ~\cite{Lattice:2015tia} & $3.72\pm0.16$\\
pQCD~\cite{Wang:2012ab}  &    $3.80{^{+0.56}_{-0.50}}$    \\
$B$-meson LCSR~\cite{Khodjamirian:2011ub}  & $3.50{^{+0.38}_{-0.33}\pm{0.11}_{\rm{exp}}}$      \\
Imsong 2014 (LCSR) ~\cite{Imsong:2014oqa} & $3.32{^{+0.26}_{-0.22}}$\\
BABAR 2012 ~\cite{Lees:2012vv} &  $3.25\pm0.31$\\
Belle 2013 ~\cite{Sibidanov:2013rkk}	& $3.52\pm0.29$\\
CKM fitter  ~\cite{Charles:2016} & $3.71{^{+0.24}_{-0.19}}$\\
UT fitter  ~\cite{Bona:2016} & $3.68\pm{0.10}$\\
FLAG 2016  ~\cite{Aoki:2016frl} & $3.62\pm0.14$\\
HFAG 2016 ~\cite{Amhis:2016xyh} & $3.67\pm0.21$\\
\hline
\end{tabular}
\caption{A comparison of $|V_{ub}|$ derived under various approaches.} \label{Vub}
\end{center}
\end{table}

A comparison of $|V_{ub}|$ under various approaches is presented in Table.~\ref{Vub}. Most of the predictions are consistent with each other within errors. Our HQEFT LCSR predictions show good agreement with the Belle measurements~\cite{Sibidanov:2013rkk}, as well as the usual LCSR predictions~\cite{Khodjamirian:2011ub, Imsong:2014oqa}.

\begin{table}[htb]
\begin{center}
\begin{tabular}{c c c }
\hline
~~~~~~~~~~~Exclusive decays~~~~~~~~~~&  ~~~~~~~~~~~~~~~~~~ \\

$B\rightarrow \pi\ell \bar{\nu}_{\ell} $ & \raisebox {2.0ex}[0pt]{${\cal B}\times10^{4}$} \\
\hline

LCSR-$\mathcal{R}$ (This work)  &  $1.41{^{+0.30}_{-0.26}}$     \\

LCSR-$\mathcal{U}$ (This work)  &  $1.39{^{+0.24}_{-0.21}}$     \\

PDG~\cite{2018pdg}  &  $1.50\pm{0.06}$     \\

BABAR 2012  ~\cite{Lees:2012vv} &  $1.44{\pm0.04}~{\pm0.06}$\\

Belle 2013 ~\cite{Sibidanov:2013rkk} & $1.49{\pm0.09}~{\pm0.07}$\\

BABAR 2010 ~\cite{delAmoSanchez:2010af} & $1.41{\pm0.05}~{\pm0.08}$\\

Belle 2010 ~\cite{Ha:2010rf} & $1.49{\pm0.04}~{\pm0.07}$\\

CLEO ~\cite{Adam:2007pv} & $1.38{\pm0.15}~{\pm0.11}$\\

HFAG 2016 ~\cite{Amhis:2016xyh}& $1.50{\pm0.02}~{\pm0.04}$\\

\hline
\end{tabular}
\caption{A comparison of the branching fraction of $B\to \pi \ell \bar \nu_\ell$ with various experimental measurements. Both LCSR-$\mathcal{U}$ and LCSR-$\mathcal{R}$ predictions are consistent with the data.}
\label{tabp1}
\end{center}
\end{table}

The correlation coefficient for the branching fractions under the LCSR-${\mathcal R}$ and LCSR-${\mathcal U}$ TFFs is, $\rho_{RU}=0.85$, which confirms the previous observation that the physical observable should be independent to the choices of correlator. More explicitly, Table.\ref{tabp1} shows a comparison of the branching fraction of $B\to \pi \ell \bar \nu_\ell$ with various experimental measurements,  which shows that both LCSR-$\mathcal{U}$ and LCSR-$\mathcal{R}$ predictions are consistent with the data.

As a final remark, it is more useful to measure the following ratio, which avoids uncertainties from the input parameters such as $|V_{ub}|$ and could be used for a precision test of SM or to find new physics beyond SM,
\begin{eqnarray}\label{Rpi}
\mathcal{R}_{\pi} &\equiv&  \frac{\mathcal{B}(B\to \pi \tau \bar{\nu_\tau})}{\mathcal{B}(B\to \pi \ell \bar{\nu_\ell})} \nonumber\\
 &=& \frac{\int^{q^2_{max}}_{m^2_{\tau}}d\Gamma(B\to \pi \tau \bar{\nu_{\tau}})/ dq^2}{\int^{q^2_{max}}_{0}d\Gamma (B\to\pi \ell \bar{\nu_\ell})/ dq^2}.
\end{eqnarray}
This ratio strongly depends on the behavior of the $B\to\pi$ TFFs. Roughly, due to the phase-space suppression, the decay width of $B\to \pi \tau \bar{\nu_\tau}$ shall be smaller than that of $B\to\pi \ell \bar{\nu_\ell}$, leading to $\mathcal{R}_{\pi}<1$ with the SM. The RBC and UKQCD collaboration predicted, $R_{\pi} = 0.69\pm{0.19}$, via the Lattice QCD calculation on the $B\to\pi$ TFFs~\cite{Flynn:2015mha}. By using our present LCSR-${\mathcal R}$ and LCSR-${\mathcal U}$ predictions for the TFFs, we obtain
\begin{eqnarray}
\mathcal{R}_{\pi}^{\cal{R}} = 0.65^{+0.13}_{-0.11}, \nonumber\\
\mathcal{R}_{\pi}^{\cal{U}} = 0.68^{+0.10}_{-0.09}.
\end{eqnarray}
These two values agree with the Lattice QCD prediction. However, the Belle measurements prefers a rather large ratio~\cite{Hamer:2015jsa}, $\mathcal{R}^{\rm{exp}}_{\pi}\simeq1.05\pm{0.51} $. A way out is to increase of the magnitude of the chiral symmetry breaking terms which are proportional to $f_0^2(q^2)$. For example, by using the type-II two Higgs doublet model (2HDM), one obtains~\cite{Bernlochner:2015mya},
\begin{eqnarray}\label{2HDM}
f_0^{2HDM}(q^2) \approx f_0^{SM}(q^2) \left(1 - \frac{\tan ^2 \beta }{m_{H^ \pm }^2}\frac{q^2}{1 - m_u/m_b}\right)
\end{eqnarray}
where the parameter $\tan \beta$ stands for the ratio of the two vacuum expectations of the Higgs doublets, and $m_{H^{\pm}}$ is the mass of the charged Higgs. If setting $\tan \beta/m_{H^+}=0.4$~\cite{Bernlochner:2015mya}, we obtain $\mathcal{R}_{\pi}^{\rm 2HDM;\cal{R}} = 0.95 ^{+0.08}_{-0.09}$ and $\mathcal{R}_{\pi}^{\rm 2HDM;\cal{U}} = 1.02 ^{+0.01}_{-0.06}$. In this ref.~\cite{Sanyal:2019xcp}, the author used  the latest $\sqrt{s}=13$ TeV CMS results to impose constraints on the charged Higgs $H^{\pm}$ parameters within the (2HDM). There are still some parameter space for the 2HDM to live. Thus we get the same as that of Ref.\cite{Bernlochner:2015mya} that there may have new physics beyond the SM. However since the Belle measurements still have large errors, we still more accurate measurements to confirm whether thus is really the case.

\section{summary}\label{Section:4}

In the paper, we have made a detailed LCSR analysis on the $B \to \pi \ell \bar\nu_\ell$ TFFs within the framework of HQEFT. As shown by Fig.(\ref{f+}), the LCSR-$\mathcal{U}$ and the LCSR-$\mathcal{R}$ for two different correlators are highly correlated with each other, e.g. $\rho_{RU} > 0.5$, especially for the $f^+(q^2)$, which provides dominant contributions to the $B\to\pi$ semileptonic decays. The LCSR method is applicable in low and intermediate $q^2$ regions and the Lattice QCD method is applicable in large $q^2$ region. After extrapolation, our HQEFT LCSR predictions on the TFFs agree well with the Lattice QCD predictions derived by HPQCD and Fermilab/MILC Collaborations. As a byproduct, by using the chiral correlator, the twist-2 LCDA provides over $97\%$ contribution to the TFFs, thus it provides a good platform for testing the properties of pion twist-2 LCDA. More over, by using the LCSR-$\mathcal{U}$ and LCSR-$\mathcal{R}$ TFFs, we obtain, $|V_{ub}|_{\cal R} = (3.45^{+0.28}_{-0.20}\pm{0.13}_{\rm{exp}})\times10^{-3}$ and $|V_{ub}|_{\cal U} = (3.38^{+0.22}_{-0.16}\pm{0.12}_{\rm{exp}})\times10^{-3}$, both of which are in accordance with experimental results within errors. The ratio $\mathcal{R}_{\pi}$ is a useful parameter to test the SM and confirm the new physics beyond the SM.

\hspace{1cm}

{\bf Acknowledgments}: We are grateful to Tao Zhong and Jun Zeng for helpful discussions. This work was supported in part by the Natural Science Foundation of China under Grant No.11375008, No.11647307, No.11625520 and No.11765007; by the Project of Guizhou Provincial Department of Science and Technology under Grant No.[2017]1089; the Key Project for Innovation Research Groups of Guizhou Provincial Department of Education under Grant No.KY[2016]028.

\appendix

\section*{Appendix: the nonlocal matrix elements and the twist-3 and 4 LCDAs}
\label{APPENDIX}

The relations between the pion LCDAs up to twist-4 accuracy and the nonlocal matrix elements are
\begin{eqnarray}
&&\langle \pi(p) |\bar u(x) \gamma_\mu \gamma_5 d(0)|0 \rangle  = - i p_\mu f_\pi \int_0^1 du e^{iup\cdot x} [ \phi_{2;\pi} (u)
\nonumber\\
&&\qquad  + x^2 g_1(u) ] + f_\pi \bigg(x_\mu - \frac{x^2 p_\mu }{p\cdot x}\bigg)\int_0^1 du e^{ - iup\cdot x} g_2 (u),\nonumber\\ \label{Eq:tw24}
\end{eqnarray}
\begin{eqnarray}
\langle \pi (p) |\bar u(x)i \gamma_5 d(0) |0\rangle  =  \frac{f_\pi m_\pi^2}{m_u + m_d}\int_0^1 {du{e^{iup \cdot x}}} {\phi_{3;\pi}^p(u)},\nonumber\\\label{Eq:tw3p}
\end{eqnarray}
\begin{eqnarray}
&&\langle \pi (p)|\bar u(x) \sigma_{\mu \nu}\gamma_5 d(0)| \rangle  =  i(p_\mu x_v - p_v x_\mu)\frac{f_\pi m_\pi^2}{6(m_u + m_d)}
\nonumber\\
&&\qquad \times\int_0^1 du e^{iup \cdot x} \phi _{3; \pi}^\sigma(u).
\end{eqnarray}
Here $f_\pi$ is the pion decay constant, $\phi_{2;\pi}(u)$, $\phi_{3;\pi}^p(u)$, $\phi _{3; \pi}^\sigma(u)$, $g_1(u)$ and $g_2(u)$ are pion twist-2, 3 and 4 LCDAs, respectively. The twist-3 LCDAs take the form
\begin{eqnarray}
\phi_{3;\pi}^p(u) &=& 1 + \frac{1}{2}B_2(3\xi^2 - 1) + \frac{1}{8}B_4(35 \xi^4 - 30\xi^2 + 3),\nonumber\\
\phi_{3;\pi}^\sigma (u) &=& 6u \bar u \bigg[ 1 + \frac{3}{2}C_2(5\xi^2 - 1) +\frac{15}{8} C_4 (21 \xi^4 - 14 \xi^2 \nonumber\\
&&  + 1)\bigg],
\end{eqnarray}
where the parameters are~\cite{Wang:2001mi, Wang:2002zba, Belyaev:1994zk, Khodjamirian:1998ji}, $B_2(\mu _b)  =  0.29$, $B_4(\mu _b) = 0.58$, $C_2(\mu _b) = 0.059$ and $C_4(\mu _b) = 0.034$. As for the twist-4 LCDAs, we adopt the ones derived by Ref.\cite{Ball:1998je}, e.g.
\begin{eqnarray}
g_1(u) &=& \frac{\bar uu}{6}[-5\bar uu(9h_{00} + 3h_{01} - 6h_{10} + 4\bar u h_{01}u+10\bar u  \nonumber\\
&& \times h_{10}u)+ a_{10}(6 + \bar uu(9 + 80\bar uu))] + a_{10}{\bar u}^3(10 \nonumber\\
&&-15\bar u+ 6{\bar u}^2)\ln \bar u + a_{10}u^3(10 - 15u + 6u^2)\ln u,\nonumber\\
g_2(u) &=& \frac{5\bar uu(u - \bar u)}{2}[4h_{00} + 8a_{10}\bar uu - h_{10}(1 + 5\bar uu) \nonumber\\
&&+ 2h_{01}(1 - \bar uu)],
\end{eqnarray}
where
\begin{eqnarray}
h_{00} & = & -\frac{\delta^2}{3}, ~  a_{10} = \delta^2\epsilon-\frac{9}{20} a^\pi_2 m_\pi^2,\nonumber\\
h_{01} & = & \frac{2}{3}\delta^2\epsilon-\frac{3}{20} a^\pi_2 m_\pi^2, ~v_{10} = \delta^2\epsilon , \nonumber\\
h_{10} & = & \frac{4}{3}\delta^2\epsilon+\frac{3}{20} a_2^\pi m_\pi^2 .\nonumber
\end{eqnarray}
where $\epsilon (\mu _b) = 0.36$, $\delta ^2(\mu _b) = 0.17\rm{GeV}^2$ ~\cite{Wang:2001mi, Wang:2002zba}.


\end{document}